\documentclass[
   ,final
  ]
  {aipproc}

\layoutstyle{6x9}

\begin{document}

\title {Tau-Mu Flavor Violation \\ and the Scale of New Physics \footnote{Talk presented by D. Black at MRST 2003 ``JoeFest'' in honor of Joe Schechter's birthday.} }

\author{Deirdre Black}{
  address={Jefferson Lab, 12000 Jefferson Ave, Newport News, VA 23606}
}

\author{Tao Han}{
  address={Department of Physics, University of Wisconsin, Madison, Wisconsin 53706}}

\author{Hong-Jian He}{
  address={Center for Particle Physics, University of Texas at Austin, Texas 78712}  }

\author{Marc Sher}{
  address={Nuclear and Particle Theory Group, College of William and Mary, Williamsburg, Virginia 23187}}



\begin{abstract}Motivated by the strong experimental evidence of large $\nu_\mu - \nu_\tau$ neutrino oscillations, we study existing constraints for related $\mu - \tau$ flavor violation.  Using a general bottom-up approach, we construct dimension-6 effective fermionic operators whose coefficients encode the scale of new physics associated with $\mu - \tau$ flavor violation, which is a piece in the puzzle of the origin of neutrino oscillations.  We survey existing experimental bounds on this scale, which arise mostly from $\tau$ and B decays.  In many cases the new physics scale is constrained to be above a few TeV.  We also discuss the operators which are either weakly constrained or, at present, subject to no experimental bounds.  
\end{abstract}

\maketitle

\section{Introduction}

In the past few years there has been a wealth of exciting results from solar, atmospheric, reactor and accelerator neutrino experiments \cite{GN03} which provide strong evidence for neutrino oscillations.  Fits to the neutrino data give large mixing angles, in contrast with the quark sector where the Cabibbo-Kobayashi-Maskawa (CKM) mixing angles are small.  Global fits \cite{GN03,PV03} give central values ${\tan}^2\theta_{\rm sol} \approx 0.46$ and ${\sin}^2 \theta_{\rm atm} \approx 0.5$ for solar and atmospheric neutrino oscillations respectively.  

The atmospheric neutrino experiments point in particular to maximal $\nu _{\mu } - \nu _{\tau }$ mixing.  In Ref. \cite{taumu} we investigated constraints on related $\mu - \tau$ flavor violation in the charged lepton sector.  Our idea was to systematically analyze bounds placed by experiment on the scale of new physics associated with a class of lepton flavor violating operators.  The goals were to idenitify in each case the kinds of process that can best be used for this purpose and to estimate the magnitude of the corresponding bounds.  This, together with other studies of lepton flavor violation, for example $\mu - e$ mixing, complements what we learn from neutrino oscillations and will help in eventually achieving an understanding of lepton flavor dynamics.  

In order to carry out a systematic study, we use a model-independent effective operator approach.  We consider an effective theory containing the Standard Model (SM) at dimension-4 and a class of dimension-6 four-fermion lepton flavor-violating operators.  A careful discussion of the construction of these operators is given in \cite{taumu}.  The effective Lagrangian is given by:
\begin{equation}
{\cal{L}}_{eff}={\cal{L}}_{SM}+\Delta {\cal{L}}_{ \mu \tau}
\end{equation}
with
\begin{equation}
\Delta {\cal{L}}_{\mu \tau }=\sum_{\alpha , \, \beta } \left[ \frac{C_{\alpha \beta }^{j}}{\Lambda^{2}}(\bar{\mu }\Gamma \tau )(\overline{{q}^{\alpha }}\Gamma q^{\beta })\right] +H.c.
\label{fourfermi}
\end{equation}
Here $\alpha$ and $\beta$ are flavor indices.  The effective interaction in Eq. (\ref{fourfermi}) paramaterizes new physics effects associated with $\mu - \tau$ flavor-violation below an ultraviolet cutoff scale, $\Lambda$.  Such dimension-6 operators could be generated, for example, from exchange of new gauge bosons or scalar (pseudoscalar) particles in the underlying theory at the scale $\Lambda$.  This is illustrated schematically in Fig. 1 and we refer the reader to \cite{taumu} for more detail.       

In Eq. (\ref{fourfermi}) $\Gamma$ denotes any of the four Dirac structures $\Gamma =1,\, \gamma _{5}\, ,\gamma ^{\mu } \, ,\gamma _{5}\gamma ^{\mu }$.  We do not consider the tensor structure here and restrict our attention to dimension-6 operators involving quark bilinears.    In the case of a linear realization of the Higgs sector the operator must be invariant under the full electroweak gauge group $SU(2)_L \times U(1)_Y$ which means that only the spinor structure $\gamma_\mu (1 \pm \gamma_5)$ is relevant.  For a non-linear realization of the gauge group, which is natural for a strongly coupled electroweak symmetry breaking sector \cite{HS02}, only $U(1)_{\rm e.m.}$ invariance is required and Eq. (\ref{fourfermi}) is the most general operator in the class we are studying.  

For a strongly coupled underlying theory with gauge coupling constant $\alpha_s = \frac{g_s^2}{4\pi} = {\cal O}(1)$ the constant  $C_{\alpha \beta }^{j}$ in Eq. (\ref{fourfermi}) is $\sim 4\pi \mathcal{O}(1)$.  We take this as a reference value throughout the present analysis, but remark that it may be scaled up or down by a factor of $4 \pi$ in a Naive Dimensional Analysis \cite{Georgi} estimation or weakly coupled scenario respectively.   Assuming that there are no accidental fine-tuned cancellations among different terms, we will consider only one operator in Eq. (\ref{fourfermi}) at a time.  

\begin{figure}
\caption{Illustration of generation of the effective four-Fermi interactions in Eq. (\ref{fourfermi}), shown in diagram (c), from exchange of (a) heavy gauge boson $Z_\mu^\prime$ or (b) new scalar or pseudoscalar particle $S^0$.}
\includegraphics[height=.27\textheight,angle=0]{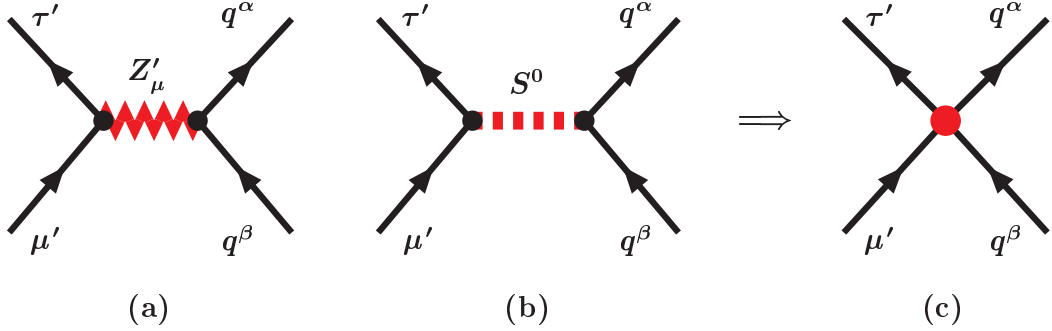}
\end{figure}
 
\section{Phenomenological Constraints}

\subsection{Light quarks: $\tau $ decays}

As explained in the Introduction, our plan is to explore experimental bounds on the scale $\Lambda$ of new physics giving rise to effective $ \mu - \tau$ flavor violating operators of the form: 
\begin{equation}
O_{ \mu \tau}=\frac{4\pi }{\Lambda ^{2}}(\bar{\mu }\Gamma \tau )(\overline{{q}^{\alpha }}\Gamma q^{\beta })
\label{operator}
\end{equation}

Experimental searches for a wide variety of lepton flavor violating (LFV) $\tau$ decay modes give strong bounds on many operators of this type.  The most recent experimental results for branching ratios for $\tau$ decay to $\mu$ and pseudoscalar or vector mesons are shown in Table 1 which comes mainly from the Particle Data Group summary \cite{PDG02}.  After our original study last year, the CLEO collaboration published \cite{CLEO02} a more stringent upper limit on the branching fraction for $\tau \rightarrow \mu K_s$ which was subsequently improved by the Belle collaboration \cite{Belle02} \footnote{The speaker thanks Jon Urheim for drawing her attention to these new results in his talk ``Rare tau decays: an experimental review'' at the CIPANP 2003 conference.}.  Here we use the latest Belle result.

\begin{table} 
\caption{Experimental upper limits on branching ratios for $\tau$ decay into pseudoscalar and vector mesons. Fourth column lists the type of operator in Eq. (\ref{operator}) which could give rise to the given LFV decay.}
\begin{tabular}{c|c|c|c}
\hline \hline
& & \\
Branching Ratio & Range of upper limits & Reference & Dirac matrix $\Gamma$ from Eq. (\ref{operator}) \\
\hline
$B(\tau \rightarrow \mu +\pi ,\eta )$ & $(4-10)10^{-6}$ & \cite{PDG02}& $\gamma_5, \, \gamma_\mu \gamma_5$ \\ 
$B(\tau \rightarrow \mu + K_s)$ & $2.7 \times {10}^{-7}$ & \cite{Belle02}& $\gamma_5, \, \gamma_\mu \gamma_5$  \\
$B(\tau \rightarrow \mu +\rho ,K^{*},\phi )$ & $(6-8)10^{-6}$ & \cite{PDG02}& $\gamma_\mu$ \\
$B(\tau \rightarrow \mu +\pi \pi ,\pi K,\pi \eta ,KK)$ & $ (7.5-22)10^{-6} $ & \cite{PDG02} & $1$  \\
& & \\
\hline\hline
\end{tabular}
\label{table1}
\end{table}

In Table 1 we have also indicated which type of Dirac structure in the effective interaction of Eq. (\ref{operator}) could give rise to each of the LFV decays listed.  In order to obtain bounds on $\Lambda$, we calculate the matrix element of the effective operator leading to each decay using vacuum insertion.  We estimate the hadronic matrix element of the quark bilinear current density using PCAC for $\Gamma = \gamma_5, \, \gamma_\mu \gamma_5$, Vector Meson Dominance for $ \Gamma = \gamma_\mu $ and the leading order Chiral Lagrangian for $\Gamma = 1$.  The resulting bounds are\footnote{In keeping with our hypothesis of considering only one operator in Eq. (\ref{fourfermi}) at a time, we treat the operators $\bar \mu \Gamma \tau \bar d \Gamma s + H.c.$ and $\bar \mu \Gamma \tau \bar s \Gamma d + H.c.$ independently and, using the new experimental limits on the branching ratio for $\tau \rightarrow \mu K_s$, quote strong bounds on each one.  However we note that if, as would be natural, these operators occur with the same coefficients then their contributions to $\tau \rightarrow \mu K_s$ would actually cancel (up to presumably small CP violating effects).} ${\cal O} (1 - 10)$ TeV and are given in Table 2.

\subsection{One b quark:  B decays}

The operators in Eq. (\ref{operator}) containing Pseudoscalar and Axial Vector bilinears and with quark-antiquark combinations $\bar b d$ and $\bar b s$ will give decays of $B$ and $B_{s}$ to $\mu \tau $.  Using vacuum insertion and a Heavy Quark Effective theory estimate $ { \left| \left< 0 \left| \bar d \gamma^5 b \right| B \right> \right| }^2 = 4m_B\beta_B^3/\pi^{3/2}$, where $\beta_B \approx 300$ MeV is a variational parameter, we obtain 
\begin{equation}
\Gamma (B\rightarrow \mu \tau )  =  \frac{\pi f_{B}^{2}m_{B}m_{\tau }^{2}}{\Lambda ^{4}}(1-\frac{m_{\tau }^{2}}{m_{B}^{2}})^{2},
\end{equation}
for $\Gamma = \gamma^\mu \gamma^5$ in Eq. (\ref{operator}), and 
\begin{equation}
\Gamma (B\rightarrow \mu \tau )  =  \frac{8m_{B}^{2}\beta _{B}^{3}}{\sqrt{\pi }\Lambda ^{4}}(1-\frac{m_{\tau }^{2}}{m_{B}^{2}})^{2},
\end{equation}
for $ \Gamma = \gamma^5 $ in Eq. (\ref{operator}).  We took $f_{B} = 200\, \mathrm{MeV}$.  The exerimental upper limits on the $\mu \tau$ decay channel are:
\begin{eqnarray*}
{\rm Br}(B^{0}\rightarrow \mu \tau ) & < & 8.3\times 10^{-4}\; \; \, [2002\; \mathrm{PDG}]\\
{\rm Br}(B_{s}\rightarrow \mu \tau ) & < & 10 \%,
\end{eqnarray*}
where the latter is a conservative estimate based on the observed $B_s$ lifetime, there being no experimental limit available yet.  The resulting bounds on $\Lambda$ are given in Table 2. 

Similarly, the operators in Eq. (\ref{operator}) containing Scalar and Vector bilinears with quark-antiquark combinations $\bar b d$ and $\bar b s$ will give the decays $B\rightarrow \pi \mu \tau ,K\mu \tau $.   The heavy-light meson matrix elements of the quark bilinear operators may be written in the form:  
\begin{equation}
<X(p_{x})|\bar{q}\gamma ^{\mu }b(0)|\bar{B}(p_{B})>\equiv f_{+}(q^{2})(p_{B}+p_{X})^{\mu }+f_{-}(q^{2})(p_{B}-p_{X})^{\mu }
\end{equation}
and we estimate the form factors $f_\pm (q^2)$ using a quark model calculation by Isgur et al (see \cite{taumu} for more detail):
\begin{equation}
f_{+}(q^{2})\approx -f_{-}(q^{2})\approx \frac{3\sqrt{2}}{8}\sqrt{\frac{m_{b}}{m_{q}}}\exp [\frac{m_{X}-E_{X}}{2\kappa ^{2}m_{X}}]
\end{equation}
where $m_{X}\approx 2m_{q},\kappa \approx 0.7$.  There is no experimental data at present on the branching ratios $B\rightarrow \pi \mu \tau , \, K\mu \tau $ so we take a conservative estimate of these branching ratios, namely ${\rm Br}(B\rightarrow \pi \mu \tau , \,K\mu \tau )<5\%$, which gives $\Lambda > 2.5$ TeV

\subsection{Top quark decays}

The effective interaction ${\cal{O}}_{\mu \tau }$ in Eq. (\ref{operator}) 
gives the following widths for the top quark decay modes $t\rightarrow c\mu \tau $ and $t\rightarrow u\mu \tau $: 
\begin{equation}
\Gamma(t\rightarrow c\mu\tau, \, u\mu\tau)  = \frac{m_{t}^{5}}{96\pi \Lambda ^{4}}
\end{equation}
for $ \Gamma = 1 \, \, {\rm and} \, \,  \gamma^5  $ in Eq. (\ref{operator}), and
\begin{equation} 
\Gamma   = \frac{m_{t}^{5}}{24\pi \Lambda ^{4}}
\end{equation}
for $ \Gamma = \gamma^\mu  \, \, {\rm and} \, \, \gamma^\mu \gamma^5 $ in Eq. (\ref{operator}), where we have neglected $m_{u,c}$.  From the CDF measurment of the ratio $\frac {{\rm Br}(t\rightarrow Wb)}{{\rm Br}(t \rightarrow Wq)}$ we make a conservative estimate that ${\rm Br}(t\rightarrow u \mu \tau, \, c\mu \tau ) <0.28$ which gives bounds on $\Lambda$ of order the top quark mass. We do not expect our effective operator approach to be consistent at such a low scale, but we note that the Naive Dimensional Analaysis estimate of the coefficient in Eq. (\ref{fourfermi}) would increase this limit by a factor of $\sqrt{4\pi }\simeq 3.5$.  Of course an improved experimental upper limit from searches for $t \rightarrow \tau \mu + {\rm jet}$ would also improve this bound.

\subsection{Other ideas for heavy quarks}

Before proceeding to look at loop-induced processes, we comment on other tree-level processes due to ${\cal O}_{\tau \mu}$ which might enable us to obtain bounds on $\Lambda$.  Firstly, the operator $\bar \mu \Gamma \tau \bar u \Gamma c$ would give $D^{0}\rightarrow \mu \tau $, but this decay is not kinematically allowed.  Secondly, $\bar \mu \Gamma \tau \bar c \Gamma c$ would give rise to charmonium ($J/\psi ,\chi _{c},\eta _{c}$) decays to $\mu \tau$, which have not been searched for yet.  Similarly, scalar and pseudoscalar $b \bar b$ decays should lead to bounds on the $\bar \mu \Gamma \tau \bar b\Gamma b$ operators which we currently have no means of constraining.

\subsection{Loop-induced processes}

We found that it is hard to constrain all operators ${\cal O}_{\mu \tau}$ with heavy quarks directly from tree-level processes.  Hence we consider one-loop
processes which induce effective light-quark vertices via insertions of the relevant heavy quark operators, for example diagrams with $W^{\pm }$ exchange as shown in Fig. 2.  On the left of this diagram we have the interaction vertex from the operator ${\cal O}_{\mu \tau}$ with a heavy quark and antiquark, labelled as up-type quarks in the Figure.  Through W exchange these are converted into a light down-type quark and antiquark.  We see that through this mechanism, if the quarks in the loop are $c$ and $\bar c$ then the final state quarks can be, for example, $s$ and $\bar s$.  This means that the operator $\bar \mu \Gamma \tau \bar c \Gamma c$ can lead to an amplitude at one loop for $\tau$ decay to $\mu$ and an $\bar s  s$ state such as the $\phi$ vector meson.  It turns out that these diagrams are non-negligible only for $\Gamma = \gamma^\mu$ or $\gamma^\mu \gamma_5$.    

\begin{figure}
\caption{Diagram relating couplings involving heavy $u$-type quark bilinears to light $d$-type quark bilinears.}
\includegraphics[height=.4\textheight,angle=0]{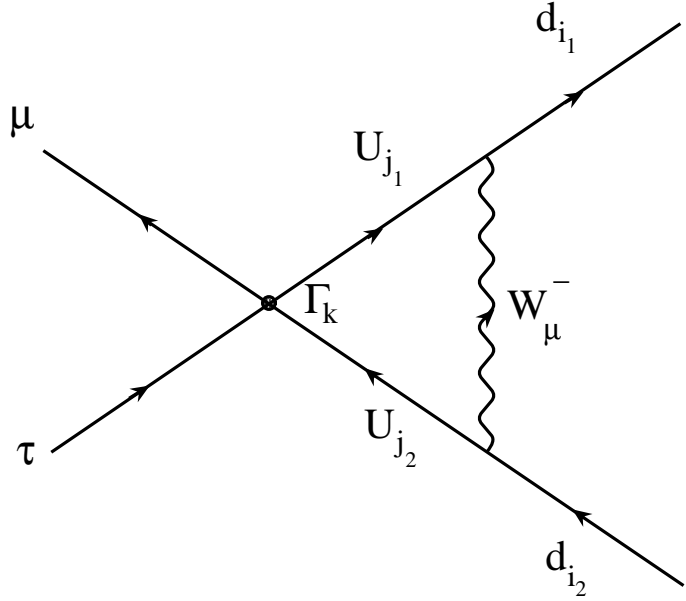}
\end{figure}

For the diagram shown in Fig. 2 we find that the operator $\bar \mu \Gamma \tau \overline {U_{j_{2}}}\Gamma U_{j_{1}}$ gives rise to a new effective interaction $(\delta C)\bar \mu \Gamma \tau \overline{d_{i_{2}}}\Gamma d_{i_{1}}$.  The loop-induced form factor, for $\Gamma = \gamma^\mu \gamma^5$ and $\gamma^\mu$ respectively, is:
\begin{equation}
\delta C_{A,V}=\frac{g^{2}V_{i_{1}j_{1}}V_{i_{2}j_{2}}^{*}}{32\pi ^{2}}(1\pm \frac{M_{j_{1}}^{2}M_{j_{2}}^{2}}{2m_{W}^{2}})\ln \frac{\Lambda }{m_{W}},
\end{equation}
where $V_{ij}$ are the CKM matrix elements associated with the W-vertices and $M_{j_{1,2}}$ are the masses of the heavy quarks in the loop.  The latter are important only for the case of the top quark.  Going back to the example given above we find that using the experimental upper limit on the branching ratio of $\tau$ to $\mu \phi$ which we listed previously, the lower bound on $\Lambda$ for the operators $\bar \mu \gamma_\mu \tau \bar c \gamma^\mu c$ and $\bar \mu \gamma_\mu \gamma^5 \tau \bar c \gamma^\mu \gamma^5 c$ is 1.1 TeV.  Other operators for which the strictest bounds come from similar one-loop processes are given in Table 2.

\section{Summary and Discussion}

We have presented a survey based on \cite{taumu} of existing constraints on flavor violation in the charged lepton sector due to effective interactions of the type $\frac{4\pi }{\Lambda ^{2}}(\bar{\mu }\Gamma \tau )(\overline{{q}^{\alpha }}\Gamma q^{\beta })$.   Most of our lower bounds on the scale of new physics responsible for these lepton flavor-violating operators come from $\tau $ and B decays at tree or one-loop level and are of order 1-10 TeV.  A complete summary of our results is given in Table 2.  

In Ref. \cite{taumu} we also considered purely leptonic effective interactions of the form $\frac{4\pi }{\Lambda ^{2}}(\bar{\mu }\Gamma \tau )(\overline{{l}^{\alpha }}\Gamma l^{\beta })$.  From experimental limits on the branching ratio for $\tau \rightarrow \mu \nu \bar{v}$ we obtained $\Lambda > 2-3$ TeV.   We find that the latest upper limits from Belle \cite{Belle02} on the branching fractions for the $\tau$ decay modes $\tau^- \rightarrow \mu^-\mu^+\mu^-, \, \mu^- \mu^- e^+, \, \mu^- \mu^+ e^- , \,\mu^- e^-e^+$ correspond to the following lower bounds on the new physics scale:  
\begin{equation}
\Lambda  > \,  19.4, \, 19.9, \, 18.5, \, 18.0 \, \, \, {\rm TeV} 
\end{equation}
and
\begin{equation}
 \Lambda  > \,  27.0, \, 28.1, \, 23.9, \, 25.5 \, \, \, {\rm TeV}
\end{equation} 
for $\Gamma = 1, \, \gamma_5 $ and $\Gamma = \gamma^\mu, \, \gamma^\mu \gamma_5 $ respectively in Eq. (\ref{operator}).

\begin{table} 
\begin{tabular}{c|cccc}
\hline \hline
& & & & \\[-2mm]
  ~~Bound~~ & $ 1 $ & $ \gamma^{~}_5$ & $\gamma_\sigma$ & 
              $\gamma_\sigma \gamma^{~}_5 $ \\
& & & & \\[-2mm]
\hline\hline
& & & & \\[-3mm]
$\bar{u} u$ & 2.6 TeV & 12 TeV & 12 TeV & 11 TeV \\           
& ~($\tau \rightarrow \mu \pi^+ \pi^-$)~ & 
~$(\tau \rightarrow \mu \pi^0$)~ & 
~($ \tau \rightarrow \mu \rho $)~ & 
~($ \tau \rightarrow \mu \pi^0 $)~  \\
& & & & \\[-3mm]
\hline
& & & & \\[-3mm]
$\bar d d$ & 2.6 TeV & 12 TeV & 12 TeV & 11 TeV \\
           & ($ \tau \rightarrow \mu \pi^+ \pi^-$) & ($\tau \rightarrow \mu \pi^0$) & ($ \tau \rightarrow \mu \rho $) & ($ \tau \rightarrow \mu \pi^0 $) \\ 
& & & & \\[-3mm]
\hline
& & & & \\[-3mm]
$\bar s s$ & 1.5 TeV & 9.9 TeV & 14 TeV & 9.5 TeV \\
           & ($ \tau \rightarrow \mu K^+ K-$) & ($\tau \rightarrow \mu \eta$) & ($ \tau \rightarrow \mu \phi $) & ($ \tau \rightarrow \mu \eta $) \\ 
& & & & \\[-3mm]
\hline
& & & & \\[-3mm]
$\bar s d$ & 2.3 TeV & 24.3 TeV  & 13 TeV & 23.6 TeV  \\
           & ($ \tau \rightarrow \mu K^+ \pi^-$) & ($\tau \rightarrow \mu K_s$) & ($ \tau \rightarrow \mu K^\star $) & ($ \tau \rightarrow \mu K_s $) \\ 
& & & & \\[-3mm]
\hline
& & & & \\[-3mm]
$\bar b d$ & 2.2 TeV  & 9.3 TeV  & 2.2 TeV  & 8.2 TeV  \\
           & ($ B \rightarrow \pi \mu \tau $) & ($B \rightarrow \mu \tau$) & ($ B \rightarrow \pi \mu \tau $) & ($ B \rightarrow \mu \tau $) \\ 
& & & & \\[-3mm]
\hline
& & & & \\[-3mm]
$\bar b s$ & 2.6 TeV  & 2.8 TeV  & 2.6 TeV  & 2.5 TeV  \\
           & ($ B \rightarrow K \mu \tau $) & ($B_{\rm s} \rightarrow \mu \tau $) & ($ B \rightarrow K \mu \tau $) & ($ B_{\rm s} \rightarrow \mu \tau $) \\ 
& & & & \\[-3mm]
\hline
& & & & \\[-3mm]
$\bar t c$ & 190 GeV  & 190 GeV  & 310 GeV & 310 GeV  \\
           & ($ t \rightarrow c \mu \tau $) & ($t \rightarrow c \mu \tau $) 
& ($B \rightarrow  \mu \tau$) & ($B \rightarrow  \mu \tau $) \\ 
& & & & \\[-3mm]
\hline
& & & & \\[-3mm]
$\bar t u$ & 190 GeV  & 190 GeV  & 650 GeV & 650 GeV  \\
           & ($t \rightarrow u \mu \tau $) & ($t \rightarrow u \mu \tau $) & ($ B \rightarrow \mu \tau $) & ($ B \rightarrow \mu \tau $) \\ 
& & & & \\[-3mm]
\hline
& & & & \\[-3mm]
$\bar c u$ & $\star$ & $\star$  & 550 GeV & 550 GeV  \\
           &  &  & ($ \tau \rightarrow \mu \phi $) & ($ \tau \rightarrow \mu \phi $) 
\\ 
& & & & \\[-3mm]
\hline
& & & & \\[-3mm]
$\bar c c$ & $\star$ & $\star$  & 1.1 TeV & 1.1 TeV  \\
           &  &  & ($ \tau \rightarrow \mu \phi $) & ($ \tau  \rightarrow \mu \phi $)
\\ 
& & & & \\[-3mm]
\hline
& & & & \\[-3mm]
$\bar b b$ & $\star$ & $\star$ & $ 180$ GeV & $\star$  \\
           &  &  &  ($\Upsilon\rightarrow \mu \tau$)  &  \\ 
& & & & \\[-3mm]
\hline
& & & & \\[-3mm]
$\bar t t$ & $\star$ & $\star$  & 75 GeV & 120 GeV  \\
           &  &  & ($ B \rightarrow \mu \tau $) & ($ B \rightarrow \mu \tau $) 
\\[1.5mm]
\hline\hline
\end{tabular}

\caption{Bounds at 90\%\,C.L. on four-Fermi flavor-violation operators 
of the form $(\bar\mu \Gamma\tau )({\bar q}^\alpha \Gamma q^\beta)$,
where $\Gamma \in ({1, \gamma_5, \gamma^\mu, \gamma^\mu \gamma_5})$. Combinations for which no 
bound has been found are marked with an asterisk,
otherwise we list the process which gives the strongest bound 
(cf. text and \cite{taumu} for details).}
\label{table2}
\end{table}

It will be interesting to further apply the bounds presented in this study to constrain specific new physics scenarios which can generate the $ \mu - \tau $ flavor violating interactions of the kind we have analyzed [c.f. Eq. (\ref{operator})].  We also note that as ongoing experimental searches for rare $\tau$ decay modes reach higher levels of precision (see for example \cite{CLEO02,Belle02}), almost half of the bounds listed in Table 2 will become more stringent.  As mentioned previously, searches for $\mu \tau$ in charmonium and bottomonium decays may help to constrain some of the operators for which we could at present find no limits.  Also we would expect searches for $B_s \rightarrow \mu \tau$ and for $B$ decays to ($\mu\tau + {\rm meson}$) to improve the values listed in Table 2 for operators involving one b quark.  It would also be interesting to look for the decay $t \rightarrow \mu \tau + {\rm jet}$ at top-quark factories such as the Tevatron Run-II and the CERN LHC.   

Finally we mention that in the section on loop-induced processes, the one-loop diagrams we considered were logarithmically divergent.  There are other important processes such as $\tau \rightarrow \mu \gamma$ which would also receive contributions at one-loop from operators of the form ${\cal O}_{ \mu \tau }$ in Eq. (\ref{operator}).  These loop diagrams are quadratically divergent (the loop consists of two quark lines) and for now we do not present bounds on the scale $\Lambda$ which would be furnished by these processes.  The identification of $\Lambda$ as the ultraviolet cutoff in these quadratically divergent loop integrals may not always be reliable \cite{London}, depending on the decoupling behavior of the specific high energy theory underlying the effective interaction ${\cal O}_{\mu  \tau}$.

\begin{theacknowledgments}
D.B. would like to thank the organizers of MRST 2003 ``JoeFest'' for a very pleasant and interesting meeting in honor and celebration of Joe Schechter's birthday.  D.B. wishes to acknowledge support from the Thomas Jefferson National Accelerator Facility
operated by the Southeastern Universities Research Association (SURA)
under DOE Contract No. DE-AC05-84ER40150.  T.H. was supported in part by a U.S. DOE grant No. DE-FG02-95ER40896 and by the Wisconsin Alumni Research Foundation.  H.-J. H. was supported by U.S. Department of Energy grant No. DE-FG03-93ER40757 and M.S. was supported by the National Science Foundation through grant PHY-9900657.

\end{theacknowledgments}

\end{document}